\documentclass[conference]{IEEEtran}

\usepackage{caption}

\usepackage{hyperref}
\usepackage{graphicx}
\usepackage{textcomp}
\usepackage{xcolor}
\usepackage[export]{adjustbox}
\usepackage{amsthm}
\usepackage{amsmath}
\usepackage{amssymb}
\usepackage{mathtools}
\usepackage{multirow}
\usepackage{pgf}
\usepackage{tikz}
\usepackage{float}
\usepackage{blindtext}
\IEEEoverridecommandlockouts
\usetikzlibrary{arrows, automata, shapes, petri, positioning, calc}

\DeclareCaptionFormat{custom}
{%
    \textbf{#1#2}\textit{\small #3}
}
\hypersetup{
	colorlinks=true,
	linkcolor={red!50!black},
	citecolor={blue!60!black},
	urlcolor={blue!80!black},
	pdfauthor={Bakullari Bianka, van der Aalst Wil M.P.},
	pdftitle={High-Level Event Mining: A Framework},
	pdfsubject={High-Level Event Mining},
	pdfkeywords={Process Mining, System-level Behavior, Congestion},
	pdfproducer={LaTeX},
	pdfcreator={pdfLaTeX},
	bookmarksopen=true
}

\newtheorem{definition}{Definition}
\def\mi#1{\mathit{#1}}

\begin{document}

\title{High-Level Event Mining: A Framework \\
\thanks{We thank the Alexander von Humboldt (AvH) Stiftung for supporting our research.}}

\author{\IEEEauthorblockN{Bianka Bakullari, Wil M.P. van der Aalst}
\IEEEauthorblockA{Chair of Process and Data Science (PADS)\\
Department of Computer Science, RWTH Aachen University\\
\texttt{\{bianka.bakullari, wvdaalst\}@pads.rwth-aachen.de}}
}

\maketitle

\begin{abstract}
Process mining methods often analyze processes in terms of the individual end-to-end process runs.
Process behavior, however, may materialize as a general state of many involved process components, which can not be captured by looking at the individual process instances.
A more holistic state of the process can be determined by looking at the events that occur close in time and share common process capacities. 
In this work, we conceptualize such behavior using high-level events and propose a new framework for detecting and logging such high-level events.
The output of our method is a new high-level event log, which collects all generated high-level events together with the newly assigned event attributes: activity, case, and timestamp.
Existing process mining techniques can then be applied on the produced high-level event log to obtain further insights.
Experiments on both simulated and real-life event data show that our method is able to automatically discover how system-level patterns such as high traffic and workload emerge, propagate and dissolve throughout the process.
\end{abstract}


\IEEEpeerreviewmaketitle

\section{Introduction} \label{sec:intro}
\subsection{Motivation}
Process mining techniques aim at getting insight into processes and improving them by using real \emph{event data} that are extracted from information systems.
Event data contain \emph{events} that occurred during process executions.
Such process executions involve various tasks, resources, facilities, costs, etc. 
Each event may concern a particular set of these process aspects.
This information is usually provided in the \emph{event attributes}.
In particular, the \emph{activity} attribute indicates the process task that was executed during the occurrence of the corresponding event. 
Each event belongs to a unique instantiation of the process, which is indicated in the \emph{case} attribute of the event.
Most process mining methods analyze the process in terms of its process instances, and claims about the process as a whole are often obtained as an aggregation of what was observed at the level of the individual cases.
E.g., process performance is assessed by aggregating the duration of individual process instances, and bottlenecks are identified by looking at the average time spent between activities.
Process behavior, however, is not necessarily a property of the individual cases as these are not isolated from each other.
Their corresponding events may demand common process capacities simultaneously.
For this reason, a more complete view on the process is necessary.
In this work, we attempt to analyze processes using a more holistic and system-aware view.
Instead of focusing on the behavior of end-to-end process runs, we provide insights over the emergence, propagation and dissolution of states that concern the system as a whole.
The state of the system is determined by the events that take place close in time.

%
%
%
\begin{figure}[h!]
\centerline{\includegraphics[scale=.38]{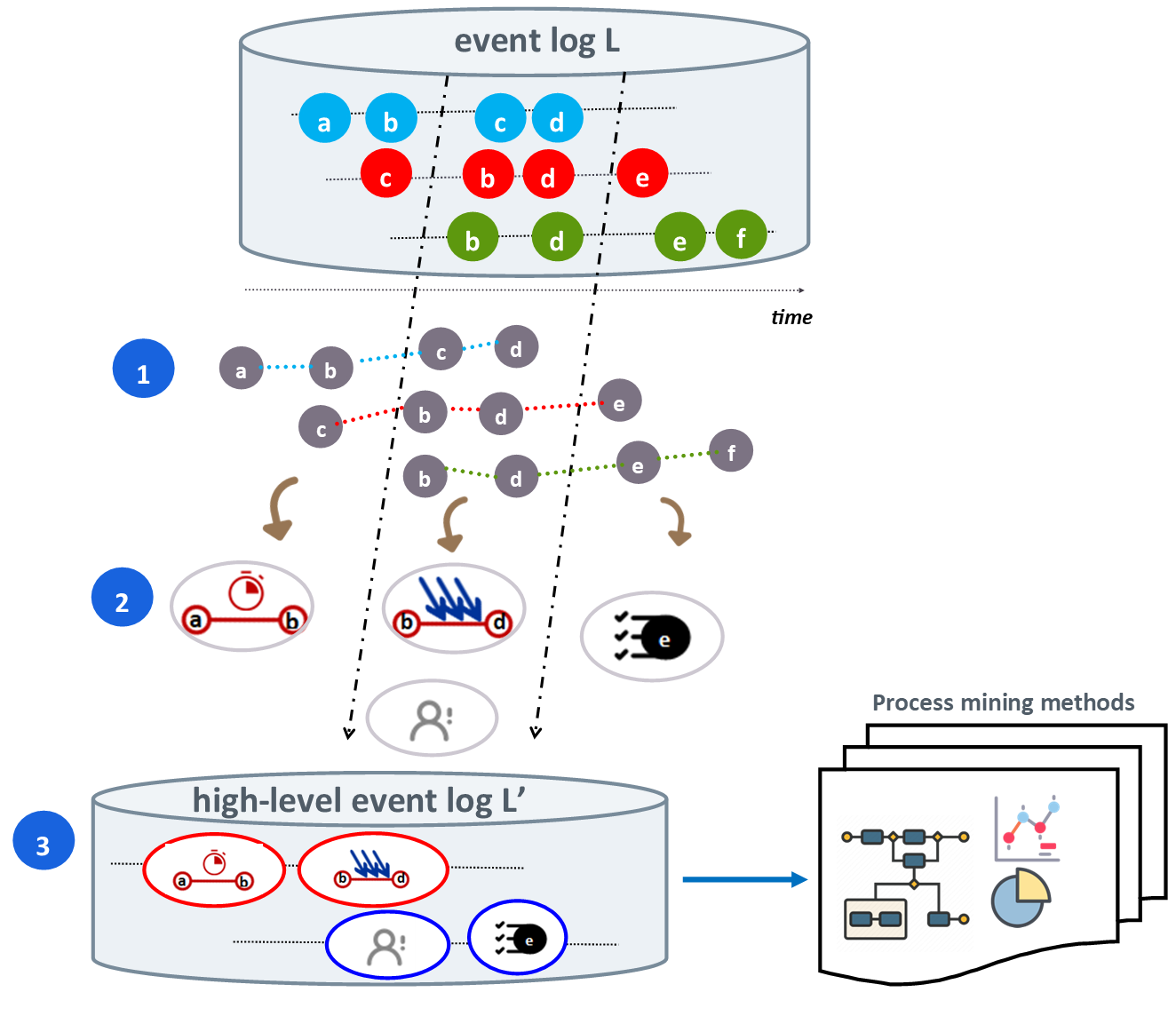}}
\caption{A visualization of the approach:
On top, the input event data showing the runs of three process instances (blue, red and green). The time scope is split into time windows (step 1), and the events within the same time window may produce certain process patterns. These patterns are captured as high-level events (step 2) which are then collected into a high-level event log (step 3).}
\label{fig: method}
\end{figure}
%
%
Fig. \ref{fig: method} provides an overview of our approach.
The input event log contains 12 events belonging to three different cases (depicted in blue, red and green).
The time scope of the process is split into time windows {(step 1)}.
The events that occur within the same window may cause emergent process behavior.
E.g., in the second window, the resource depicted in gray is overloaded with work and there are many cases that recently finished activity $b$ and are waiting for activity $d$ to happen.
We explicitly capture such behavior in the form of \emph{high-level events} {(step 2)}, which similarly to ``normal" events, describe what happened and when it happened in the process.
The generated high-level events are collected into the so-called \emph{high-level event log} {(step 3)} where a new case perspective is introduced (the output log in Fig. \ref{fig: method} contains two cases depicted in red and blue).
The event log format enables applying existing process mining techniques to gain insights over the captured high-level process behavior.
\subsection{Example}
\begin{figure}[h]
\centerline{\includegraphics[scale=.4]{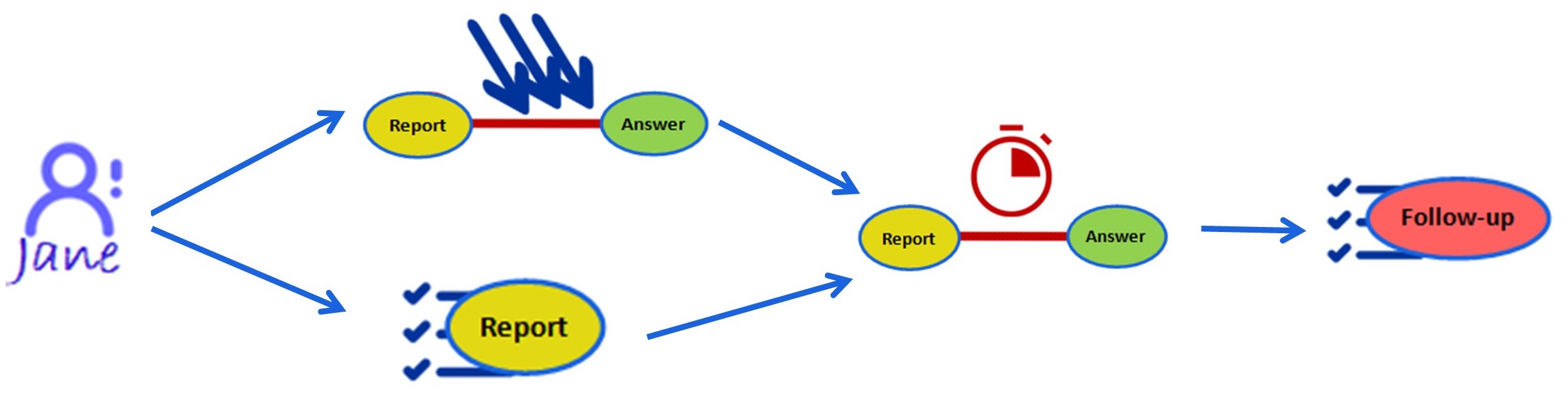}}
\caption{A directed graph visualizing the process patterns behind the undesired behavior explained in the example:
When Jane has high workload (violet resource icon), she tends to execute many ``Report" tasks (indicated by the ticks beside activity ``Report").
Thus, multiple requests accumulate in-between tasks ``Report" and ``Answer" (depicted with the blue arrows).
This causes high waiting times in this segment (red timer icon), which triggers more ``Follow-up" activities (indicated by the ticks beside activity ``Follow-up").}
\label{fig: jane}
\end{figure}

Consider the following scenario: The service desk of a traveling agency handles questions and requests by their customers. 
For each request, a report is filed and an answer is sent to the customer.
If the answer is delayed, customers may send a follow-up question asking about the status of their request.
These questions require the same attention that new requests do and they cause additional traffic in the overall process.
When many such questions are received, the added workload delays the responses even further, leading to even more follow-up questions.
Jane is one of the resources who answers customers' requests.
Whenever she sees requests piling up, she prefers to first finish all the paperwork by filing the reports and only answers the customers afterwards.
While waiting for a reply, these customers tend to become impatient and send a follow-up question.
The agency is interested in discovering whether there are certain patterns in the process executions that often correlate with high numbers of follow-up questions.
A helpful insight to the agency would be to see how Jane's workload, her workflow, and the number of follow-up requests are related to each-other (Fig. \ref{fig: jane}).

%
%
%
The introduced example points out that process behavior can have the following properties:
1) it may concern different parts of the process (activities, resources, states between two activities) which are not properties of the cases themselves, 
2) it is not necessarily persistent; it can emerge and dissolve at different time periods, and
3) it is not isolated; it may be both the consequence and cause of other process behavior.
\subsection{Approach}
In our work, we assume we are provided with an \emph{event log}, which is a collection of recorded events and their attributes.
We address each of the aforementioned points in a dedicated step:
First, we define new features which capture both the kind of detected behavior and the activity, resource, or process location it affects.
We also refer to the latter as the underlying \emph{component} of the feature.
Second, we determine non-overlapping time-windows and partition the events into the time windows within which they occurred.
Every feature of interest is then evaluated over each time-window.
Each outlier measurement generates a dedicated \emph{high-level event}.
Third, any pair of components underlying the features is assigned a proximity value that reflects how ``close" those components are in terms of process executions. 
E.g., a pair of activities that are always executed subsequently are close, whereas any resource is distant from the activities it never executes.
A high-level event \emph{propagates} to another high-level event in the subsequent time window whenever their underlying components are close enough. 
Any maximal sequence of high-level events connected to each other through propagation forms a \emph{cascade}.
4) We log all high-level events and generate a new event log, which we call a \emph{high-level event log}.
For any high-level event, the activity attribute reflects the feature for which the outlier value was measured, the timestamp corresponds to the time window in which the high-level event emerged, whereas the case attribute coincides with the cascade the high-level event belongs to.

The configurable parts of the framework include the choice of the features and the thresholds which determine the emergence and correlation of the high-level events.
%
%
\section{Related Work}\label{sec:rel}
\emph{Performance analysis techniques} typically concern the time perspective, aiming to obtain insights regarding duration, bottlenecks, and delays \cite{perf-literature-review}.
An initial approach demonstrating how time information is aggregated from all end-to-end process runs is shown in \cite{vdaalst-ts}.
When a new case is still running, a prediction for the remaining time is estimated from the remaining times of past cases similar to the current one, and any interplay between different cases is not considered.
More refined methods compute delays and waiting times separately for cases that occur within different contexts \cite{context-bart}. 
These approaches, however, view the effect of any particular context on performance as a static property of the process which is satisfied by all or most cases happening within that context.
Using the \emph{performance spectrum} \cite{perf-spectrum}, one can detect arising system-level behavior such as high loads, delays or batching \cite{batch-eva} using visual analysis.
This technique can visualize the load and waiting times only for the cases that run through a particular pair of activities, while dependencies between the performance patterns concerning different activity pairs may not be visible.
In \cite{dumas-stages}, the authors assume each case in the process goes through a sequence of stages, and they visualize how queues, incoming- and outgoing rates in each stage develop over time.
In contrast, we do not make any assumption about the control-flow of the process, and any outlier measurement regarding performance is conceptualized and logged as an event of its own.
The authors in \cite{zahra} were the first to propose a method that explicitly captures undesired system-level behavior such as blockages and delays in the form of system-level events.
Each system-level event is characterized by the process location and the outlier behavior (high load or long waiting time) measured in that location during a particular time window.
Events that are immediately close both in time and location are connected into sequences called \emph{cascades}.
The most frequent cascades reveal how undesired system-level behavior arises and propagates throughout the process.
We take this approach further by introducing a framework within which these system-level events are naturally incorporated and extended.
They are conceptualized as a specific type of high-level events that concern congestion features.
In addition, we capture resource behavior in terms of workload and task execution, and also define a new proximity function which exploits the control-flow of the process to determine ``how" close the underlying process components are.
This proximity value is then used to decide whether two high-level events should be connected into a cascade.
Similar work extending the idea from \cite{zahra} was done in \cite{anna} where DBSCAN is used to find frequent sequences of anomalies arising as system-level behavior.
In our approach, we instead collect the generated high-level events and log them into a new event log, where the cascade membership is incorporated in the case identifier.
This way, one can exploit the rich body of existing process mining techniques to analyze the higher-level behavior.
The authors in \cite{queue-weidlich} recognize queueing as a consequence of inter-case dependencies that may lead to delays in the process.
They predict these delays for running cases using techniques referred to as \emph{queue mining}.
In our work, large accumulations of process instances waiting to be served are captured using dedicated types of high-level events, and our focus is not on predicting the remaining time for a specific case, but rather discovering the development and dependencies among the high-level events themselves.
The work by van Zelst et al. in \cite{abstraction} provides a good overview over \emph{event abstraction techniques} in process mining.
These techniques mainly aim to alter the granularity level of the provided event data by bringing it closer to the level of detail that enables correct and understandable process insights.
The authors in \cite{Mahsa} proposed an approach for defining aggregated process variables and calculating them over time steps.
The calculated measurable aspects of processes, also known as coarse-grained process logs, present event logs at a higher level of granularity than event logs.
These process logs are used for higher-level simulation and prediction of process states.
Our generated high-level events are not coarser representations of the input event data, but rather events describing new  emergent behavior caused by the combination of ``low-level events"—none of which displays the behavior individually.
\emph{Concept drift detection techniques} \cite{concept-drift} aim at identifying process changes that happen while the process is being analyzed.
In our framework, significant process changes may become apparent as they cause the type of generated high-level events to also vary over time.
%
%
%
%
\section{Preliminaries}\label{sec:prelim}
In the remainder, for any set $X$, set $\mathcal{P}(X)$ denotes the power set of $X$, $\mathcal{B}(X)$ denotes the set of all multisets over set $X$, and $X^*$ denotes all sequences over set $X$.
\begin{definition}[Events, Event attributes]\label{def:events}
    $\mathcal{U}_{\mi{ev}}$ is the universe of events,
    $\mathcal{U}_{\mi{act}}$ is the universe of activities, 
    $\mathcal{U}_{\mi{case}}$ is the universe of cases, 
    $\mathcal{U}_{\mi{time}}$ is the universe of timestamps, and 
    $\mathcal{U}_{\mi{res}}$ is the universe of resources.
    $\mathcal{U}_{\mi{val}}$ is the universe of values, and $\mathcal{U}_{\mi{map}}=\mathcal{U}_{\mi{att}} \not \rightarrow \mathcal{U}_{\mi{val}}$ is the universe of event attribute-value mappings.
    We assume that $\{\mi{act}, \mi{case}, \mi{time}, \mi{res}\} \subseteq \mathcal{U}_{\mi{att}}$ and $\mathcal{U}_{\mi{act}}$, $\mathcal{U}_{\mi{case}}$, $\mathcal{U}_{\mi{time}}$, $\mathcal{U}_{\mi{res}} \subseteq \mathcal{U}_{\mi{val}}$.
\end{definition}

\begin{definition}[Event log]\label{def:log}
An \emph{event log} is a tuple $L=(E,\pi)$ where $E \subseteq \mathcal{U}_{\mi{ev}}$ is the set of events and $\pi \in E \rightarrow \mathcal{U}_{\mi{map}}$ such that for any $e \in E$:
$\{\mi{act}, \mi{case}, \mi{time}, \mi{res}\} \allowbreak \subseteq \mi{dom}(\pi(e))$ and
$\pi(e)(\mi{case}) \in \mathcal{U}_{\mi{case}}$ is the case of $e$, $\pi(e)(\mi{act}) \in \mathcal{U}_{\mi{act}}$ is the activity of $e$, 
$\pi(e)(\mi{time}) \in \mathcal{U}_{\mi{time}}$ is the timestamp of $e$, and
$\pi(e)(\mi{res}) \in \mathcal{U}_{\mi{res}}$ is the resource of $e$.
\end{definition}

For any $x \in \mathcal{U}_{\mi{att}}$ and any event $e \in E$ of a given log $L=(E,\pi)$, we write $\pi_{x}(e)$ instead of $\pi(e)(x)$.

\begin{definition}[Steps]\label{def:steps}
Given an event log $L=(E,\pi)$, the set $\mi{steps}(L)=\{(e_1,e_2) \in E \times E \mid \pi_{\mi{case}}(e_1) = \pi_{\mi{case}}(e_2) \ \wedge \ \pi_{\mi{time}}(e_1) < \pi_{\mi{time}}(e_2) \ \wedge \ \forall_{e \in E \setminus \{e_1, e_2\}} \ \pi_{\mi{case}}(e) = \pi_{\mi{case}}(e_1) \Rightarrow \pi_{\mi{time}}(e) \leq \pi_{\mi{time}}(e_1) \ \vee \ \pi_{\mi{time}}(e) \geq \pi_{\mi{time}}(e_2) \}$ contains the \emph{steps} of event log $L$.
Moreover, for any $(e_1, e_2) \in \mi{steps}(L)$, we say that event $e_1$ \emph{triggers} $e_2$.
Conversely, $e_2$ \emph{is triggered by} $e_1$.
\end{definition}
Two events of a given log constitute a step if they belong to the same case and no other event of that case occurred in-between.
These pairs are equivalent to the so-called ``directly-follows" event pairs of a log.
\begin{definition}[Framing]\label{def:framing}
A \emph{framing} is a non-decreasing function ${\phi \in \mathcal{U}_{\mi{time}} \rightarrow \mathbb{N}}$.
For any $w \in \mi{rng}(\phi)$, $\lfloor w \rfloor = \mi{min} \{t \in \mathcal{U}_{\mi{time}} \mid \phi(t) = w \}$ and $\lceil w \rceil = \mi{max} \{t \in \mathcal{U}_{\mi{time}} \mid \phi(t) = w \}$ denote the minimal and maximal timestamps assigned to $w$.
Given an event log $L$ and a framing $\phi$, the set $W_{L,\phi}=\{w \in \mathbb{N} \mid 
{\mi{min}\{\phi(\pi_{\mi{time}}(e)) \mid e \in E\}} 
\leq w \leq 
{\mi{max}\{\phi(\pi_{\mi{time}}(e)) \mid {e \in E}\}} \}$ 
is called the \emph{window set} of event log $L$ w.r.t. $\phi$.
Each $w \in W_{L,\phi}$ is called a window and for any $e \in E$, we say \emph{$e$ occurred during $w$} whenever $\lfloor w \rfloor \leq \pi_{\mi{time}}(e) < \lceil w \rceil$.
\end{definition}
%
%
%
%
\section{Method}\label{sec:method}
\subsection{Defining high-level events}\label{sub:hle}
The choice of the high-level features determines the type of the high-level events that can emerge in the process.
Each feature is evaluated at every time window, and outlier measurements are captured in the form of high-level events. 
\begin{definition}[Feature evaluation]\label{def:eval}
$\mathcal{U}_{\mi{hlf}}$ is the universe of high-level feature names.
Given an event log $L$ and framing $\phi$, the \emph{feature evaluation} of $L$ at any window $w \in W_{L,\phi}$ is a partial function $\mi{eval}(L,w) \in \mathcal{U}_{\mi{hlf}} \not \rightarrow \mathbb{R}$ which maps high-level features to real numbers.
\end{definition}
For any $\mi{hlf} \in \mathcal{U}_{\mi{hlf}}$, for simplicity we write $\mi{eval}_{\mi{hlf}}(L,w)$ instead of $\mi{eval}(L,w)(\mi{hlf})$.
\begin{definition}[High-level events]\label{def:hle}
Let $L$ be an event log and $\phi$ a framing.
Let $\mi{HLF} \subseteq \mathcal{U}_{\mi{hlf}}$ be a set of high-level features and $\mi{thresh} \in \mi{HLF} \rightarrow \mathbb{R}$ a function mapping each feature onto a corresponding threshold. 
The set $\mathcal{H}_{L,\phi,\mi{HLF}, \mi{thresh}} = \{(\mi{hlf}, w)\in \mi{HLF} \times W_{L,\phi} \mid \mi{eval}_{\mi{hlf}}(L,w) \geq \mi{thresh}(\mi{hlf})\}$ contains all \emph{high-level events} obtained from log $L$, framing $\phi$, and feature set $\mi{HLF}$ with threshold function $\mi{thresh}$. 
I.e., a feature-window pair $(\mi{hlf},w)$ generates a high-level event whenever the value assigned to $\mi{hlf}$ at window $w$ is higher than or equal to the corresponding threshold.
\end{definition}
As congestion is a universal property of processes that emerges on a system-level, we instantiate our framework by defining high-level features that are related to congestion.
In the following, we explain the motivation behind the choice of those high-level features.

\begin{figure}[h]
\centerline{\includegraphics[scale=.2]{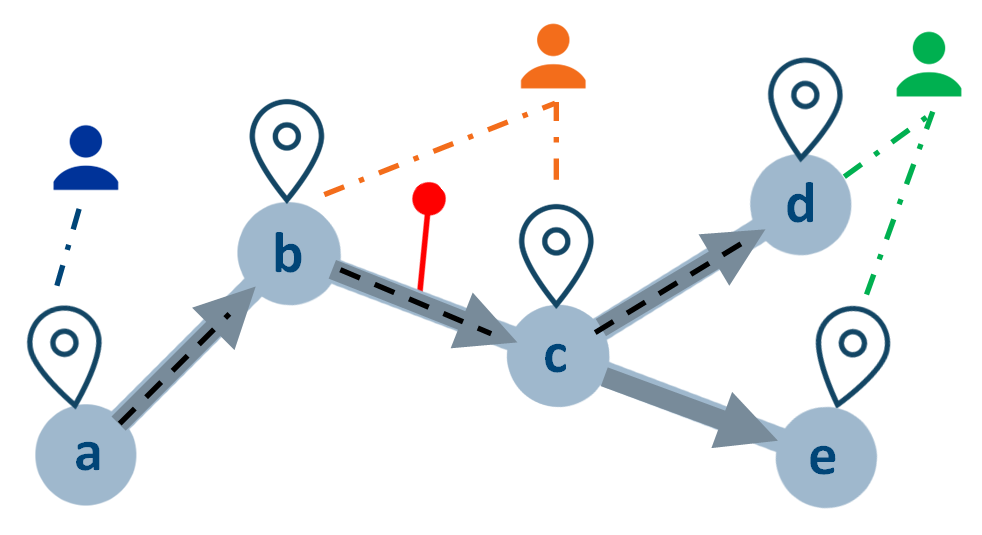}}
\caption{Some process visualized as a road map with activities depicted as landmarks. 
Each activity has a corresponding responsible resource.
There are two possible runs through this process: one executing activities $\langle a,b,c,d \rangle$, and one executing activities $\langle a,b,c,e \rangle$.
The path of some running case is $\langle a,b,c,d \rangle$ (the black dashed line).
Currently, this case occupies segment $(b,c)$ (the red pin).}
\label{fig: roadmap}
\end{figure}

For each case in a process, there is a sequence of events describing its process run.
Consider the process visualized in Fig. \ref{fig: roadmap}.
Suppose that the event sequence $\langle e_a, e_b, e_c, e_d \rangle$ executing activities $\langle a,b,c,d \rangle$ was recorded during the run of some case in this process.
At any moment between the first and last event of the sequence, we can determine the last previous event and the first next event of the case.
Assume that at a given time, $e_b$ was the last recorded event.
Thus, the next event to occur is $e_c$.
In terms of congestion, the case is ``located" between activities $b$ and $c$. 
At any timestamp $t$ between $\pi_{\mi{time}}(e_b)$ and $\pi_{\mi{time}}(e_c)$, this case influences congestion at segment $(b,c)$ by increasing the load by one and the current waiting time by $t-\pi_{\mi{time}}(e_b)$. 
In the mean time, this indicates a new task piling up for the resource that will execute the next activity (here $c$).
Whenever a resource executes an activity, the load in the process is shifted from the previous segment to the next segment.
In this example, as soon as $c$ occurs, the case moves on to occupy the next segment (which for this case is $(c,d)$).
This continues until the case is complete.

%
%
%
%
%
At any time interval throughout the process, each activity may be executed multiple times in the context of several different cases.
Moreover, each resource has a specific set of new tasks piling up and/or being executed within that time interval.
How long each case has to wait for the next activity to occur likely depends on the current workload of the responsible resource.
The process run of any case is both influencing and influenced by the other process instances which have overlapping life cycles and share common resources or paths through the process. 
We refer to activities, resources, and process segments as the congestion components of any process.
\begin{definition}[Activities, Resources, Segments]
Given an event log $L=(E,\pi)$, $A(L):=\{\pi_{\mi{act}}(e) \mid e \in E\}$ is the \emph{activity set} of $L$, $R(L):=\{\pi_{\mi{res}}(e) \mid e \in E\}$ is the \emph{resource set} of $L$, and 
$S(L):=\{(\pi_{\mi{act}}(e_1),\pi_{\mi{act}}(e_2)) \mid (e_1,e_2) \in \mi{steps}(L)\}$ is the \emph{segment set} of $L$.
The segment set of an event log consists of all activity pairs that correspond to a step in the log.
Additionally, for any $a \in A(L)$, we define $E{\upharpoonright}_a := \{e \in E \mid \pi_{\mi{act}}(e) = a\}$ as the \emph{$a$-events} of $L$, for any $r \in R(L)$, we define $E{\upharpoonright}_r := \{e \in E \mid \pi_{\mi{res}}(e) = r\}$ as the \emph{$r$-events} of $L$, and
for any $s=(a_1,a_2) \in S(L)$, we define $\mi{steps}(L){\upharpoonright}_s := \{(e,e') \in \mi{steps}(L) \mid \pi_{\mi{act}}(e) = a_1\ \wedge \ \pi_{\mi{act}}(e') = a_2\}$ as the \emph{$s$-steps} of $L$.
\end{definition}
%
%
Next, we define some high-level features which relate to congestion patterns on the activity-, resource-, and segment-level.
For each of those features, we describe how the value $\mi{eval}$ of that feature is computed given an event log and a time window.

Suppose we are given an event log $L$ and window set $W_{L,\phi}$ w.r.t. some framing $\phi$.
For any $a \in A(L)$, feature ${\mi{exec}\text{-}a \in \mathcal{U}_{\mi{hlf}}}$ stands for the number of executed $a$-events, and for any $w \in W_{L,\phi}$: $\mi{eval}_{\mi{exec}\text{-}a}(L,w) = |\{ e \in E{\upharpoonright}_a  \mid \lfloor w \rfloor \leq \pi_{\mi{time}}(e) < \lceil w \rceil \}|$.
Similarly, for any $r \in R(L)$, the feature ${\mi{do}\text{-}r \in \mathcal{U}_{\mi{hlf}}}$ stands for the number of $r$-events, and for any ${w \in W_{L,\phi}}$: $\mi{eval}_{\mi{do}\text{-}r}(L,w) = |\{ e \in E{\upharpoonright}_r \mid \lfloor w \rfloor \leq \pi_{\mi{time}}(e) < \lceil w \rceil \}|$.
For any step $(e,e') \in \mi{steps}(L)$, the execution of $e$ triggers a new task (namely $\pi_{\mi{act}}(e')$) for resource $\pi_{\mi{res}}(e')$.
For any $r \in R(L)$, we capture such new workload using feature $\mi{todo}\text{-}r$.
For any $w \in W_{L,\phi}$, $\mi{eval}_{\mi{todo}\text{-}r}(L,w) = |\{ e' \in E{\upharpoonright}_r \mid  \exists_{e \in E} \ (e,e') \in \mi{steps}(L) \ \wedge \ \lfloor w \rfloor \leq \pi_{\mi{time}}(e) < \lceil w \rceil \}|$ is the number of $r$-events that are triggered during $w$.
The total workload of any resource $r \in R(L)$ during some window is the sum of all $r$-events that either occur or are waiting to be handled by $r$ at that time window.
We capture the total workload of any $r \in R(L)$ using feature $\mi{wl}\text{-}r$, where for any $w \in W_{L,\phi}$:
$\mi{eval}_{\mi{wl}\text{-}r}(L,w) = |\{e' \in E{\upharpoonright}_r \mid \lfloor w \rfloor \leq \pi_{\mi{time}}(e') < \lceil w \rceil  \ \vee \ \exists_{e \in E} \ (e,e') \in \mi{steps}(L) \ \wedge \ \pi_{\mi{time}}(e) < \lceil w \rceil \ \wedge \ \pi_{\mi{time}}(e') > \lfloor w \rfloor\}|$.
The congestion in a particular segment of the log can be captured by e.g., looking at the number of steps that enter, exit or cross the segment during a particular window.
For any segment $s=(a_1, a_2) \in S(L)$, let $\mi{arrive}(s,w)=\{(e,e') \in \mi{steps}(L){\upharpoonright}_s \ \mid \ \lfloor w \rfloor \leq \pi_{\mi{time}}(e) < \lceil w \rceil \}$ be the set of steps that arrive at $s$ during $w$, let $\mi{leave}(s,w)=\{(e,e') \in \mi{steps}(L){\upharpoonright}_s \ \mid \ \lfloor w \rfloor \leq \pi_{\mi{time}}(e') < \lceil w \rceil\}$ be the set of steps that leave $s$ during $w$, 
and $\mi{cross}(s,w)=\{(e,e') \in \mi{steps}(L){\upharpoonright}_s \ \mid \ \pi_{\mi{time}}(e) < \lceil w \rceil \ \wedge \  \pi_{\mi{time}}(e') \geq  \lfloor w \rfloor \}$ the set of steps that cross $s$ during $w$.
Note that $\mi{arrive}(s,w)$ and $\mi{leave}(s,w)$ are subsets of $\mi{cross}(s,w)$.
For any $s \in S(L)$, the number of steps entering, exiting or being in progress at $s$ at any window $w \in W_{L,\phi}$ is indicated by the values $\mi{eval}(L,w)$ assigns to features $\mi{enter}\text{-}s$, $\mi{exit}\text{-}s$ and $\mi{progr}\text{-}s$ respectively.
Here, we have
$\mi{eval}_{\mi{enter}\text{-}s}(L,w) = |\mi{arrive}(s,w)|$,
$\mi{eval}_{\mi{exit}\text{-}s}(L,w) = |\mi{leave}(s,w)|$, and
$\mi{eval}_{\mi{progr}\text{-}s}(L,w) = |\mi{cross}(s,w)|$.
Lastly, we introduce feature $\mi{delay}\text{-}s$ for any $s \in S(L)$ which stands for the delay at segment $s$.
Its value is computed as the average waiting time that has accumulated from all the steps in progress at $s$ during a given time window.
More precisely, for any $w \in W_{L,\phi}$:
$
    \mi{eval}_{\mi{delay}\text{-}s}(L,w) = \frac{1}{\mi{eval}_{\mi{progr}\text{-}s}(L,w)} \cdot \\ \Big( \sum_{(e,e') \in \mi{leave}(s,w)} \ \pi_{\mi{time}}(e') - \pi_{\mi{time}}(e) + \\ \sum_{(e,e') \in \mi{cross}(s,w) \setminus \mi{leave}(s,\mi{tw})} \ \lceil w \rceil - \pi_{\mi{time}}(e) \Big).
$

\vspace*{0.1cm}
Each of these congestion features describes congestion in terms of a particular \emph{view} (e.g., number of executions, workload, delay) on a particular \emph{component} (activities, resources, segments).
To determine the threshold of any of these features, we can consider the multiset of values that $\mi{eval}$ assigns to all features of the same view across all windows.
Using some $p \in [0,1]$, the value of $\mi{thresh}$ can be the smallest value of this multiset that is equal or greater than the $p*100$-th percentile.
For example, given event log $L$ and framing $\phi$, the delay at some segment $s \in S(L)$ at window $w \in W_{L,\phi}$ generates a high-level event if it is high enough compared to all delays measured over all segments across the windows $W_{L,\phi}$.
If $p = 0.9$, then only the $10\%$ highest measured delays will generate high-level events, whereas if $p=0.7$, the $30\%$ highest measured delays will generate high-level events.
\subsection{Correlating high-level events}\label{sub:corr}
\begin{definition}[Proximity, Propagation]\label{def:proximity}
Let $\mathcal{H}_{L,\phi,\mi{HLF}, \mi{thresh}}$ be the set of high-level events obtained from event log $L$, framing $\phi$, and feature set $\mi{HLF}$ with threshold function $\mi{thresh}$. 
For any two high-level events $\mi{hle}_1, \mi{hle}_2 \in \mathcal{H}_{L,\phi,\mi{HLF}, \mi{thresh}}$, ${\bowtie_L}(\mi{hle}_1,\mi{hle}_2) \in [0,1]$ yields the \emph{proximity} between $\mi{hle}_1$ and $\mi{hle}_2$ with 0 being the farthest and 1 being the closest.
Moreover, for some ${\lambda \in [0,1]}$, there is a \emph{propagation} from $\mi{hle}_1$ to $\mi{hle}_2$ w.r.t. $\lambda$ (denoted ${\mi{hle}_1 \leadsto_{\lambda} \mi{hle}_2}$) if and only if ${\bowtie_L}(\mi{hle}_1, \mi{hle}_2) \geq \lambda$.
I.e., one high-level event propagates to another if and only if they are closer to each other than some threshold $\lambda$.
\end{definition}
The proximity values describe how close any pair of high-level events are.
The higher the proximity value, the higher the chance that the pair of high-level events will be considered as correlated.
When two high-level events are close enough w.r.t. some threshold, we say that the earlier event \emph{propagated} to the later one.
Note that by the definition of the high-level events, how ``proximity" between any two high-level events is measured can depend both on the underlying features, as well as on the time windows that generated them.
\begin{definition}[Cascade]\label{def:casc}
Let $\mathcal{H}_{L,\phi,\mi{HLF}, \mi{thresh}}$ be the set of high-level events generated from event log $L$, framing $\phi$, and feature set $\mi{HLF}$ with threshold function $\mi{thresh}$.
Given some $\lambda \in [0,1]$, any function $\mi{casc}_{\lambda} \in {\mathcal{H}_{L,\phi,\mi{HLF}, \mi{thresh}} \rightarrow \mathbb{N}}$ is called a \emph{cascade identifier} function if and only if for any two high-level events $\mi{hle}, \mi{hle}' \in \mathcal{H}_{L,\phi,\mi{HLF}, \mi{thresh}}$, the following holds:
\begin{align*}
\mi{casc}_{\lambda}(\mi{hle}) = \mi{casc}_{\lambda}(\mi{hle}') \ \Leftrightarrow \ 
\mi{hle} \leadsto_{\lambda} \mi{hle}' \ \vee \\
\exists_{\mi{hle}_1,...,\mi{hle}_n \in \mathcal{H}_{L,\phi,\mi{HLF}, \mi{thresh}}} \ \mi{hle} \leadsto_{\lambda} \mi{hle}_1 \ \wedge \ \mi{hle}_n \leadsto_{\lambda} \mi{hle}' \ \wedge \\
 \forall_{1 \leq i < n} \ \mi{hle}_i \leadsto_{\lambda} \mi{hle}_{i+1}.
\end{align*}
That is, a cascade identifier assigns two high-level events the same number if and only if there is either a direct or indirect propagation from one high-level event to the other.
\end{definition}
In the following, we propose a way for computing proximity values between all high-level event pairs whose underlying features correspond to those introduced in the last subsection (\ref{sub:hle}).
When correlating high-level events that emerge from the congestion-related features introduced previously, we abstract from the view of the feature (e.g., delay, workload, etc.) and only evaluate proximity based on the underlying component (activity, resource or segment).
On one hand, this way we naturally incorporate the knowledge about the control-flow and work distribution that is present in the log.
On the other hand, we avoid making assumptions on how the features correlate to each other, as this is in fact part of the high-level process behavior that we wish to discover.
Moreover, for simplicity, we assume propagation only occurs between subsequent time windows and thus, assign positive proximity values only to high-level events of adjacent time windows.
This way, any two high-level events from non-adjacent windows can only be correlated indirectly through a chain of propagation.
The high-level events we generate all have a process component they relate to: an activity, a resource or a segment.
In terms of congestion, all these components are related---resources execute the activities, the execution of an activity ``moves" the corresponding case from one segment to another, and new work piles up for the resource that must handle the next activity.
Traffic that arises at some time window may persist across multiple time windows that follow, and moreover, it can trigger more traffic in its ``neighborhood".
For any pair of components, we determine a $\mi{link}$ value which reveals how closely connected these components are based on the provided event data.
The proximity value of any pair of congestion-related high-level events that emerge in adjacent time windows is then equal to the $\mi{link}$ value of their underlying components.
We reuse the road map metaphor from Fig. \ref{fig: roadmap} to motivate how we determine the link values.
Each activity requires a responsible resource and its execution either increases or decreases the load from its adjacent segments (e.g. executions of $b$ from Fig. \ref{fig: roadmap} affect the orange resource and segments $(a,b)$ and $(b,c)$).
Resources immediately affect the load shifts between the segments whose underlying activities they are responsible for executing (e.g. when the orange resource from Fig. \ref{fig: roadmap} executes $c$, it removes load from $(b,c)$ and shifts it towards $(c,d)$ or $(c,e)$, depending on the path of the current case).
Resources also influence each other's workload if they execute neighboring activities (e.g. whenever the orange resource from Fig. \ref{fig: roadmap} executes $c$, a new task piles up for the green resource who is responsible for activities $d$ and $e$). 
The $\mi{link}$ values reflect how connected component pairs are by exploiting the frequency of these dependencies.

Next, we show how one can compute the $\mi{link}$ values and thus, the proximity values for high-level event pairs.
Let $L=(E,\pi)$ be an event log.
For any two components $x_1, x_2 \in A(L) \cup R(L) \cup S(L)$ with $x_1 \neq x_2$, $\mi{link}_L(\{x_1,x_2\}) \in [0,1]$ shows how connected $x_1$ and $x_2$ are in the process with 0 being the farthest and 1 being the closest.
If $x_1, x_2 \in A(L)$ or $x_1, x_2 \in R(L)$, then
$\mi{link}_L(\{x_1, x_2\}) = 
\mi{max}\{ 
\lvert \mi{steps}(L){\upharpoonright}_{(x_1,x_2)}\rvert / 
\lvert E{\upharpoonright}_{x_1}\rvert , \allowbreak
\lvert \mi{steps}(L){\upharpoonright}_{(x_2,x_1)}\rvert / 
\lvert E{\upharpoonright}_{x_2}\rvert
\}$,
where $\mi{steps}(L){\upharpoonright}_{(r_1,r_2)} = \{(e,e') \in \mi{steps}(L) \mid \pi_{\mi{res}}(e) = r_1\ \wedge \ \pi_{\mi{res}}(e') = r_2\}$ for any resource pair $r_1, r_2 \in R(L)$.
I.e., activity pairs and resource pairs are closer the more often they correspond to events that directly follow each-other.
For any $a \in A(L)$ and $r \in R(L)$,
$\mi{link}_L(\{a, r\}) = 
\mi{max} \{
\lvert E{\upharpoonright}_{a} \cap E{\upharpoonright}_{r}\rvert / 
\lvert E{\upharpoonright}_{a}\rvert , \ \allowbreak
\lvert E{\upharpoonright}_{a}  \cap  E{\upharpoonright}_{r}\rvert / 
\lvert E{\upharpoonright}_{r}\rvert
\}$.
I.e., a resource-activity pair is closer the more often it is that resource executing the activity, or it is that activity being executed when the resource is working.
For any $a \in A(L)$ and $s=(a_1, a_2) \in S(L)$, if $a \in \{a_1, a_2\}$ then
$\mi{link}_L(\{a, s\}) = 
\mi{max} \{
\lvert \mi{steps}(L){\upharpoonright}_{(a_1, a_2)}\rvert, 
\lvert \mi{steps}(L){\upharpoonright}_{(a_2, a_1)}\rvert
\} /
\lvert E{\upharpoonright}_{a}\rvert$. 
Else, $\mi{link}_L(\{a, s\}) = 0$.
In other words, each activity can only be close to its adjacent segments.
The $\mi{link}_L$ value of any activity-segment pair reflects how often the occurrence of the activity indicates a case that is entering or exiting the segment.
For any $r \in R(L)$ and $s=(a_1, a_2) \in S(L)$, let $\mi{steps}(L){\upharpoonright}_{s,r} := 
\{(e,e') \in \mi{steps}(L){\upharpoonright}_{s} \mid 
e \in E{\upharpoonright}_{r} \ \vee \ 
e' \in E{\upharpoonright}_{r}\}$.
Then, $\mi{link}_L(\{r, s\}) = 
\mi{max} \{
\lvert \mi{steps}(L){\upharpoonright}_{s,r} \rvert / 
\lvert E{\upharpoonright}_{r} \rvert,
\lvert \mi{steps}(L){\upharpoonright}_{s,r} \rvert /
\lvert \mi{steps}(L){\upharpoonright}_{s} \rvert
\}$.
I.e., a resource-segment pair is closer the more often the resource executes one of the activities of the segment.
Lastly, for any $a_1, a_2, a_3 \in A(L)$ such that $s=(a_1, a_2), s'=(a_2, a_3) \in S(L)$, let $\mi{steps}(L){\upharpoonright}_{(a_1,a_2,a_3)} := 
\{(e_1,e_2,e_3) \in E \times E \times E \mid 
(e_1,e_2) \in \mi{steps}(L){\upharpoonright}_{s} \ \wedge \
(e_2,e_3) \in \mi{steps}(L){\upharpoonright}_{s'}\}$.
Then, $\mi{link}_L(\{s, s'\}) = 
\mi{max} \{
\lvert \mi{steps}(L){\upharpoonright}_{(a_1,a_2,a_3)} \rvert / 
\lvert \mi{steps}(L){\upharpoonright}_{s}, \linebreak
\lvert \mi{steps}(L){\upharpoonright}_{(a_1,a_2,a_3)} \rvert /
\lvert \mi{steps}(L){\upharpoonright}_{s'} \rvert
\}$.
Otherwise, for any $s=(a_1,a_2), s'=(a_1', a_2') \in S(L)$ such that $\{a_1, a_2\} \cap \{a_1', a_2'\} = \emptyset$, $\mi{link}_L(\{s, s'\}) = 0$.
I.e., only segments that share a common activity can be close.
The $\mi{link}_L$ value of any such pair of segments reflects the fraction of cases which cross both segments from those that cross at least one of the segments.
%
%
%
%
%
%
%
\subsection{Generating a high-level event log}
\begin{definition}[High-level event log]\label{def:hlel}
Let $\mathcal{H}_{L,\phi,\mi{HLF},\mi{thresh}}$ be the set of high-level events generated from log $L$, framing $\phi$, and feature set $\mi{HLF}$ with threshold function $\mi{thresh}$. 
For some $\lambda \in [0,1]$, the \emph{high-level event log} corresponding to $\mathcal{H}_{L,\phi,\mi{HLF}, \mi{thresh}}$ is a new event log ${L'=(E',\pi')}$ such that $E'= \{e_{\mi{hlf},w} \mid (\mi{hlf},w) \in \mathcal{H}_{L,\phi,\mi{HLF},\mi{thresh}} \}$ and
$\{\mi{act},\mi{case},\mi{time}\} \subseteq \mi{dom}(\pi'(e))$ for all $e \in E'$.
For any $e_{\mi{hlf},w} \in E'$: $\pi_{\mi{act}}(e_{\mi{hlf},w}) = \mi{hlf}$,
$\pi_{\mi{time}}(e_{\mi{hlf},w}) = \lfloor w \rfloor$, and
$\pi_{\mi{case}}(e_{\mi{hlf},w}) = \mi{casc}_{\lambda}((\mi{hlf},w))$.
\end{definition}
There is a one-to-one correspondence between the high-level events and the events in the generated high-level event log.
The activity attribute (which we refer to as the \emph{high-level activity}) describes the high-level feature, the timestamp indicates the time window in which the high-level event emerged, whereas the case identifier is determined by the cascade it belongs to.
Note that high-level events that emerge within the same time window are assigned the same timestamp.
Thus, events of the same case in the high-level event log may be partially ordered.
%
\section{Evaluation}\label{sec:eval}
The method is available as a Python implementation\footnote{\url{https://github.com/biankabakullari/hlem-framework}}.
It can automatically generate a high-level event log from any given event log.
The user can determine the set of features and components they wish to focus on, and select the feature thresholds and $\lambda$.
In our evaluation, we use \emph{Cortado} \cite{cortado} to visualize the most frequent variants in the high-level event log, as this tool can handle partially ordered event data.
By using a fixed total order between all generated high-level activities, all partially-ordered cascades can be flattened into sequences of totally ordered high-level events. 
We use the flattened log to obtain a Directly-Follows Graph (DFG) that displays the typical high-level activity sequences in the newly created event log.
\begin{table*}[t]
\centering
\begin{tabular}{c|c|c|c|c|c|c|c|c|c|c}
\multicolumn{3}{c}{} & \multicolumn{2}{c}{\textbf{enter-(report, answer)}} & \multicolumn{2}{c}{\textbf{delay-(report, answer)}} & \multicolumn{2}{c}{\textbf{workload-Jane}} & \multicolumn{2}{c}{\textbf{exec-follow}} \\
Week & \# events & \# hle & count & avg. value (visits) & count & avg. value (hours) & count & avg. value (tasks) & count & avg. value (executions) \\ \hline
1   & 2451  & 4     & 4     & 3.25  & -     & -     & -     & -     & -     & -     \\ \hline
2   & 17608 & 1624  & 250   & 3,77  & 439   & 1,41  & 340   & 36    & 404   & 9.29  \\ \hline
3   & 9408  & 724   & 118   & 3.75  & 191   & 1,50  & 146   & 39    & 175   & 9.79  \\ \hline
4   & 8750  & 191   & 80    & 3.46  & -     & -     & -     & -     & 80    & 5.05   \\ \hline
5   & 5791  & 102   & 31    & 3.13  & -     & -     & -     & -     & 63    & 6.19  \\ \hline
6   & 13016 & 1150  & 190   & 3.84  & 288   & 1.44  & 227   & 37    & 300   & 8.97  \\ \hline
7   & 2577  & 251   & 39    & 3.64  & 72    & 1.34  & 48    & 33    & 59    & 9.47 \\ \hline
\end{tabular}
\caption{A summary of the data appearing in the high-level event log obtained from the process described in Section \ref{sec:intro}.
The columns show for each week, the number of events in the initial log, the number of high-level events, and the counts (i.e. absolute amounts) and average values for the most frequent high-level activities. Empty entries should be interpreted as zeros. }\label{table: simulation}
\end{table*}
\subsection{Simulated log}
\begin{figure}[h!]
\centerline{\includegraphics[scale=.3]{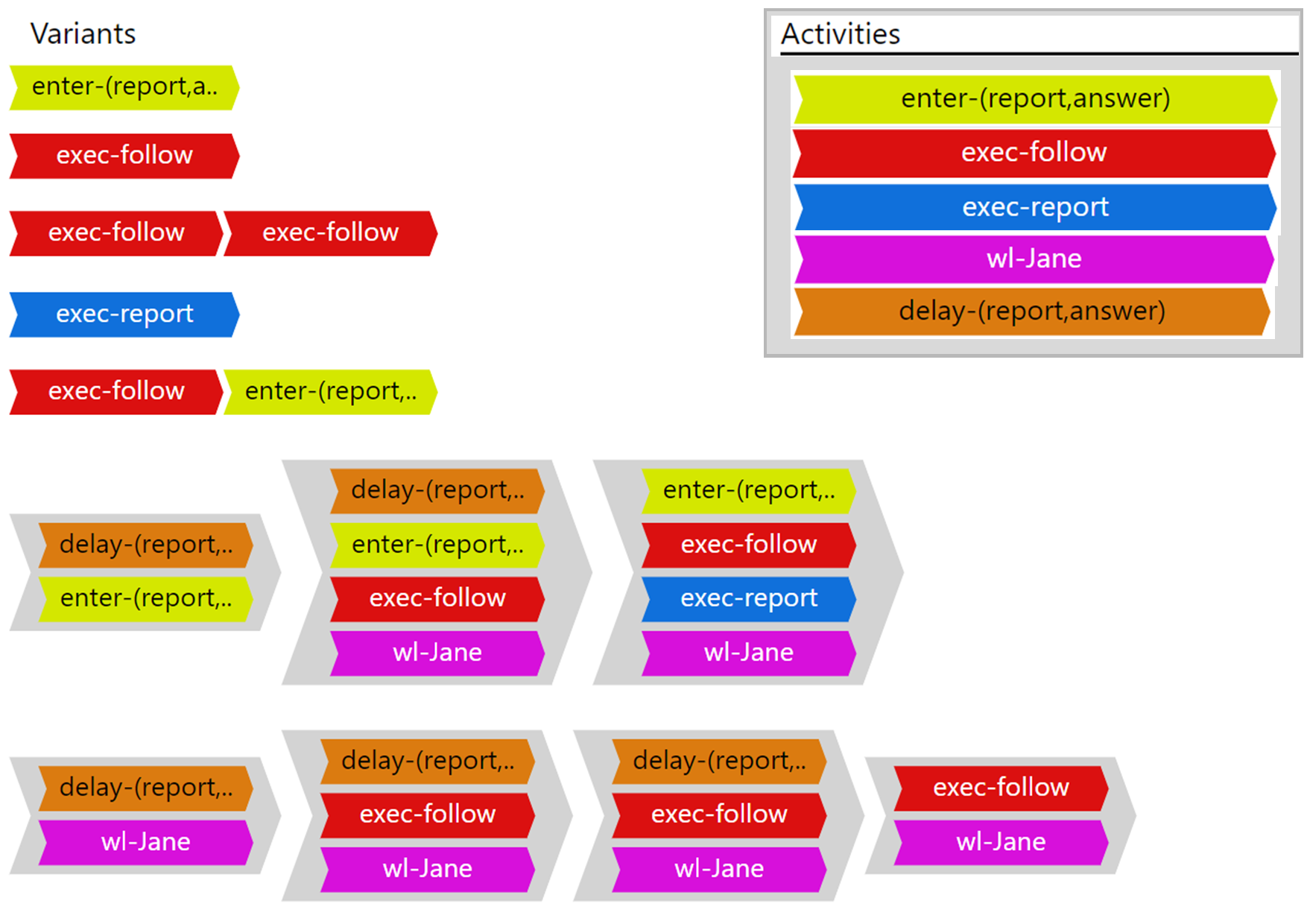}}
\caption{The seven most frequent variants of the high-level cases visualized with \emph{Cortado} \cite{cortado}.
Each colored chevron shows a high-level activity, and activities depicted within the same gray chevron occurred at the same time window.
The two most frequent variants show that activities \emph{enter-(report,answer)} and \emph{exec-follow} often emerge for a single time window, followed by a third variant where \emph{exec-follow} emerges in two subsequent windows.
The variants with partially ordered activities show how the workload of Jane often emerges simultaneously with a high entering load and long delay in the segment \emph{(report, answer)}. Afterwards, \emph{exec-follow} also appears in the cascade.}
\label{fig: cortado_delay}
\end{figure}

\begin{figure}[h!]
\centerline{\includegraphics[scale=.34]{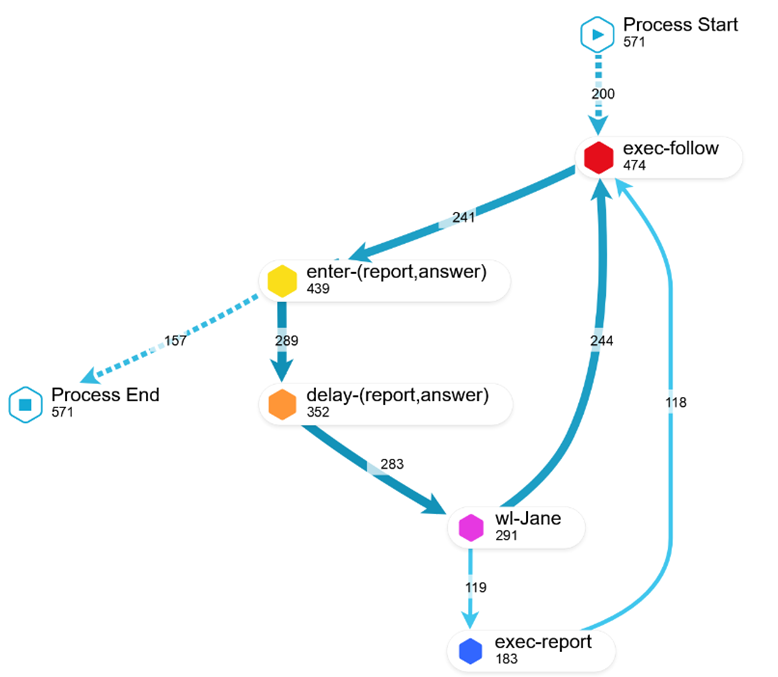}}
\caption{The DFG discovered from the flattened high-level event log using \emph{Celonis}.
The four high-level activities \emph{wl-Jane}, \emph{exec-follow}, \emph{enter-(report, answer)}, and \emph{delay-(report, answer)},  which were shown to happen simultaneously in the partially ordered log, appear in a cycle in the DFG of the flattened log.
The similar frequencies of the arcs connecting those activities, reveal that the same cycle appears over and over in the process.}
\label{fig: celonis_delay}
\end{figure}

We evaluate our method on a simulated log that displays the behavior of the process we described in the example in Section \ref{sec:intro}.
The log was simulated using \emph{CPN Tools} \cite{cpn} and is available on GitHub.
Each case in the event log represents a customer request.
For each request there is a report that has to be filed (activity \emph{report}) and the customer must receive a reply (activity \emph{answer}).
If the reply takes too long, customers may send a follow-up question (activity \emph{follow}).
Jane is one of the resources who answers customers' requests.
When Jane has a high workload, she first files all the reports of the queueing requests and only answers the customers afterwards.
This behavior increases the probability of a waiting customer to send a follow-up question.
In our log, new requests arrive at different rates across 7 weeks (see Table \ref{table: simulation}).
During weeks 1,4,5, and 7, a new case arrives every 10--15 minutes.
Weeks 2,3, and 6 are busier with a new case arriving every 3--5 minutes.
Assuming all resources are available and no follow-up question is sent, each case can take between 4--10 minutes to complete.
Some customers wait at least three hours before sending a follow-up question, whereas others wait no longer than one hour.
Table \ref{table: simulation} summarizes the number of initial events, high-level events, high-level activities and their average values for each week.
Whenever there are many high-level events related to Jane's high workload, the numbers of cases entering and being delayed in the segment between \emph{report} and \emph{answer} are also higher (see weeks 2, 3 and 6).
This demonstrates how the two concurrent activities \emph{report} and \emph{answer} are mostly executed in this order when Jane is busy.
The busy weeks 2, 3 and 6 are also reflected in the number of occurrences of activity \emph{follow}.
When looking at the absolute amounts of the generated high-level events (the ``count" columns), it is obvious that they are particularly high (low) when the number of initial events is high (low).
This variance in the process traffic is, however, invisible when looking at the average values among the weeks.
The only exception is the number of follow-up requests whose average values still reflect that variance.
This demonstrates how (un)desired behavior can be clearly visible in a process (similar to the follow-up requests), but the patterns that lead to that behavior may disappear if one looks at the general picture which considers all cases within a broad time scope.
High-level events manage to capture process behavior that is dynamic, possibly short in its lifespan, but with significant consequences for the process.
Fig. \ref{fig: cortado_delay} displays the discovered cascades as partially ordered sequences of high-level activities.
We also flattened the log and used \emph{Celonis} to discover the DFG in Fig. \ref{fig: celonis_delay}.
In both figures, the high-level activities are colored according to the legend shown in Fig. \ref{fig: cortado_delay}.
%
\subsection{Real-life event log}
\begin{figure}[h!]
\centerline{\includegraphics[scale=.28]{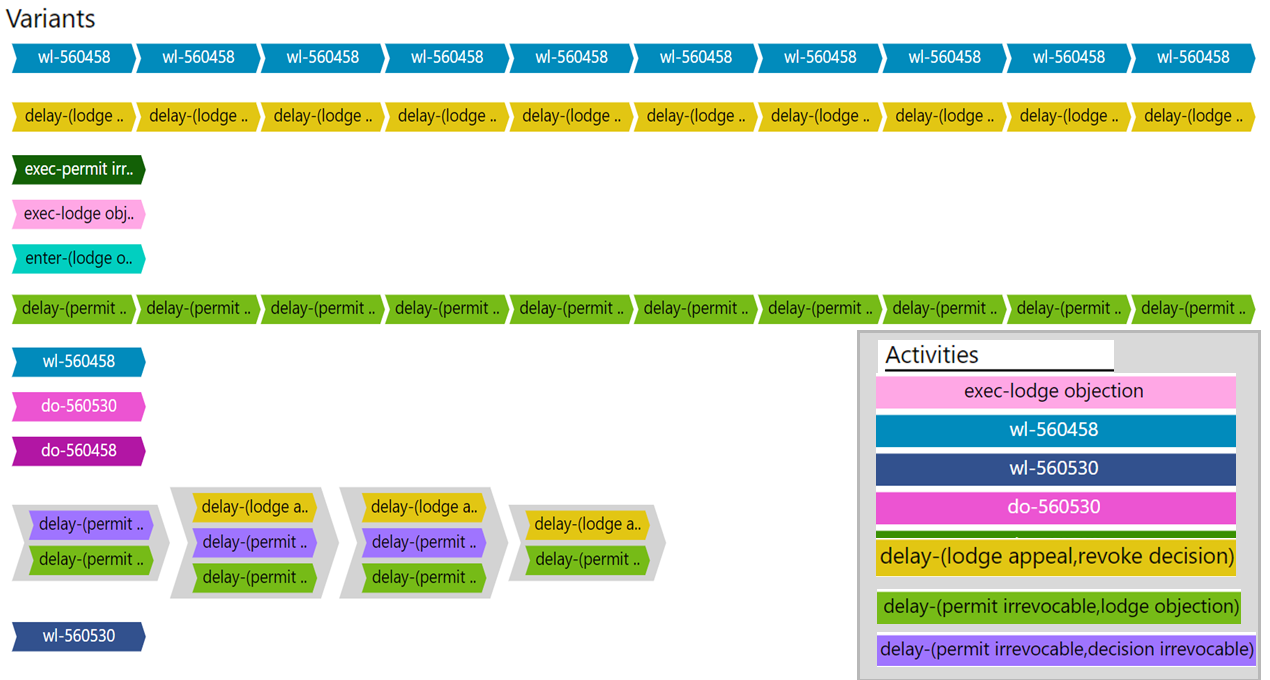}}
\caption{The most frequent variants of the high-level events obtained from the ‘objections and complaints’ subprocess of municipality 2 in BPI Challenge 2015.}
\label{fig: bpic2}
\end{figure}

\begin{figure}[h!]
\centerline{\includegraphics[scale=.28]{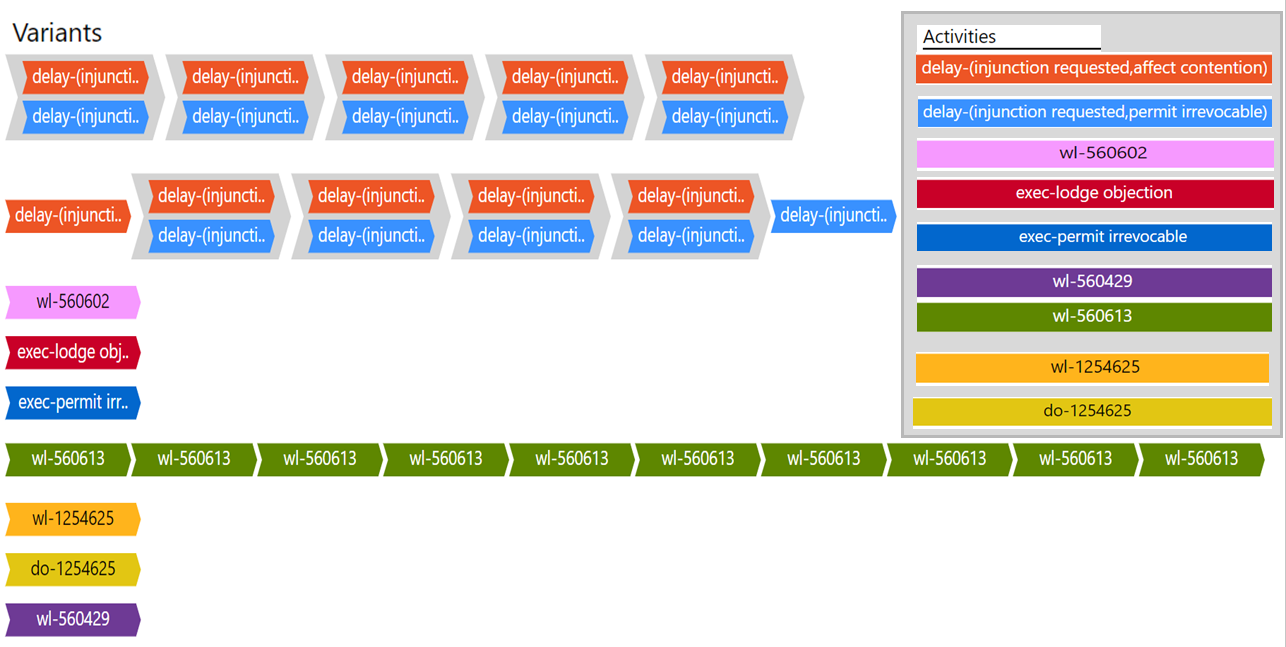}}
\caption{The most frequent variants of the high-level events obtained from the ‘objections and complaints’ subprocess of municipality 5 in BPI Challenge 2015.}
\label{fig: bpic5}
\end{figure}

We also applied our method on the real-life event log \emph{BPI Challenge 2015}\footnote{\url{10.4121/uuid:31a308ef-c844-48da-948c-305d167a0ec1}}.
The data are provided by five Dutch municipalities and concern building permit applications.
Here, we analyze the ‘objections and complaints’ subprocess recorded for municipalities 2 and 5.
The most frequent variants are depicted in Fig. \ref{fig: bpic2} and \ref{fig: bpic5}.
Even though the process and the typical activities are the same among the two municipalities, the results reveal that congestion arises in different process parts.
For municipality 2 (Fig. \ref{fig: bpic2}), the phase between activities \emph{lodge appeal} and \emph{revoke decision} seems to have long delays (yellow) most often.
Also, many cases enter the phase identified with activity \emph{permission irrevocable} (dark green).
There are often simultaneous persisting delays between this phase and the phase identified with activity \emph{decision irrevocable} (violet) or activity \emph{lodge objection} (green) which follow. 
For municipality 5 (Fig. \ref{fig: bpic5}), there are often persisting delays between activity \emph{injunction requested} and the activities \emph{affect contention} (orange) and \emph{permit irrevocable} (blue).
In both municipalities, the involved resources often underlie workload-related high-level events: e.g., resources 560458 (blue), 560530 (pink) in municipality 2 and 560613 (green), 560602 (pink), and 560429 (violet) in municipality 5.
The presence of these high-level events shows that the workload for many resources varies strongly throughout the process.
In municipality 5, resources with id 560429 and 560613 are often affected simultaneously by a high workload, which might indicate common responsibilities or high handover of work between them.
%
%
%
%
%
\section{Conclusion and Future Work}\label{sec:conc}
In this work, we introduced a new framework for detecting, logging and interpreting process behavior that can not be captured in terms of individual process runs.
Such behavior emerges on the system level, it arises and dissolves at different time periods, and it may trigger other, similar behavior.
We conceptualized this behavior in terms of high-level events, which similar to the ``normal" events indicate what happened and when it happened.
We used congestion as a universal trait of processes, to demonstrate what a high-level event can describe and how it can be detected from the event data.
Moreover, we showed how one can exploit the control-flow information present in the data to correlate the detected high-level events.
We evaluated the approach both on simulated data and on real-life event data.
The results show that the method is able to discover high-level patterns that may be invisible when using traditional case-oriented techniques.

This work can be extended in many directions.
On a more technical side, the method can be improved to be more robust to the parameter configuration.
Finding a good setting able to detect most interesting patterns on the right level of detail requires good knowledge of the process at hand.
Considering only directly-follows steps to determine the position of a case in the overall ``process map" has the following advantages: 
On one hand, every case can only occupy one segment simultaneously.
Thus, the estimated load for segments and resources simply reflects the number of cases, making it easy to interpret.
On the other hand, this also enables considering the stage between concurrent activities as a location in itself with its own load and waiting time.
This way, we were able to detect the particular resource behavior in the simulated log.
However, if the concurrent activities are independent from each other, the high-level events they produce should not be correlated despite their consecutive occurrences.
Another challenge is to visualize such data in a more intuitive and interactive way.
On the conceptual side, high-level behavior in processes does not have to be related to congestion.
Moreover, high-level behavior concerning activities and other process entities may not necessarily need a (unique) case notion present in the data.
Lastly, the method should be further evaluated on real life processes prone to congestion and costly errors.
%
%
%
%
\bibliographystyle{IEEEtran}
\bibliography{bibliography-short.bib}
\end{document}